%% file: main.tex
\documentclass{sig-alternate-ipsn13}
\usepackage{amsmath}
\usepackage{amsfonts}
\usepackage{color}
\usepackage{graphicx}

\newtheorem{rem}{Remark}

\input{Ahmed_defs.tex}

\begin{document}

\title{Deep Value of Information Estimators for Collaborative Human-Machine Information Gathering\titlenote{This material is based upon work partially supported by the National Science Foundation under Grant No. CNS-1464279 and the IUCRC Center for Unmanned Aerial Systems (CUAS).}}
%
%
%
%
%

\numberofauthors{5} 
%
\author{
%
%
\alignauthor Kin Gwn Lore\\
       \affaddr{Iowa State University}\\
       \affaddr{Ames, IA-50010, USA}\\
       \email{kglore@iastate.edu}
\alignauthor Nicholas Sweet\\
       \affaddr{University of Colorado}\\
       \affaddr{Boulder, CO-80309}\\
       \email{nisw6751@colorado.edu}
\alignauthor Kundan Kumar\\
       \affaddr{Iowa State University}\\
       \affaddr{Ames, IA-50010, USA}\\
       \email{kkumar@iastate.edu}
\and  
\alignauthor Nisar Ahmed\\
       \affaddr{University of Colorado}\\
       \affaddr{Boulder, CO-80309}\\
       \email{nisar.ahmed@colorado.edu}
\alignauthor Soumik Sarkar\\
       \affaddr{Iowa State University}\\
       \affaddr{Ames, IA-50010, USA}\\
       \email{soumiks@iastate.edu}
}



\maketitle
\begin{abstract}
Effective human-machine collaboration can significantly improve many learning and planning strategies for information gathering via fusion of `hard' and `soft' data originating from machine and human sensors, respectively. However, gathering the most informative data from human sensors without task overloading remains a critical technical challenge. In this context, Value of Information (VOI) is a crucial decision-theoretic metric for scheduling interaction with human sensors.
We present a new Deep Learning based VOI estimation framework that can be used to schedule collaborative human-machine sensing with computationally efficient online inference and minimal policy hand-tuning. Supervised learning is used to train deep convolutional neural networks (CNNs) to extract hierarchical features from `images' of belief spaces obtained via data fusion. These features can be associated with soft data query choices to reliably compute VOI for human interaction. The CNN framework is described in detail, and a performance comparison to a feature-based POMDP scheduling policy is provided. The practical feasibility of our method is also demonstrated on a mobile robotic search problem with language-based semantic human sensor inputs.
\end{abstract}

\section{Introduction}\label{sec:intro}
The notion of human-machine interaction for cooperative problem solving has attracted much interest in various cyber-physical system domains. Much of the human-machine interaction literature in the AI, robotics and controls communities tend to focus on the idea of humans acting in the role of collaborative planners or controllers for machine counterparts. In this work, however, we examine the problem of using humans as `soft data sensors' for intelligent machine systems. In particular, we focus on the question of how human observations can be used to augment state estimates of measurable dynamical physical states that must be monitored continuously by conventional `hard' sensor data used by machines (object position, velocity, attitude, temperature, size, mass, etc.).
%

Several modeling approaches have been developed to exploit soft human sensing across a variety of interfaces, e.g. verbally reported range and bearing for target localization \cite{Kaupp-JFR-2007}; verbal and sketch-based detection/no detection reports for target search \cite{Bourgault08, Ahmed-ICCPS-2015}; and semantic language inputs for target localization \cite{AhmedTRO13}. While these works have largely focused on developing human sensor models and suitable data fusion algorithms for blending hard and soft data, relatively little work has been done on \emph{active soft sensing}, i.e., intelligent querying of human sensors to gather information that would be most beneficial for complex machine planning and/or perception tasks. Active sensing problems have a rich tradition in target tracking and controls communities, but have focused on hard data sources such as radar, lidar, cameras, etc. One particularly relevant issue is the \textit{sensor scheduling problem}, i.e., the selection of most appropriate sensing assets given constraints on how many can be tasked to deliver data any given instant (e.g., due to bandwidth or computational limits). In this work, we address the problem of scheduling interactions between human sensors and their machine counterparts: what information should be solicited from human sensors to get most valuable soft information back out for machine counterparts, and when/how should such soft information be obtained? These issues can be tackled within formal planning frameworks that seek to maximize the \emph{value of information} (VOI) under uncertainty, so that soft data is only provided by human sensors if it is worth the cost of using limited machine and human cognitive resources to obtain it. However, exact inference and optimization for VOI-based human-machine interaction is computationally expensive, and approximations for efficient online interaction must be sought.  

This paper proposes a new deep learning based VOI estimation approach that aims to learn the policy reward given a joint state-action configuration. Deep learning is an emerging branch of machine learning that uses multiple levels of abstractions (from low-level features to higher-order representations, i.e., features of features) learnt from data without any hand-tuning.
Among various deep learning tools, convolutional neural network (CNN)~\cite{KSB10a} is an attractive option for extracting pertinent features from images in a hierarchical manner for detection, classification, and prediction. In addition to this deep learning architecture, other architectures such as deep nelief networks, deep autoencoders, and deep recurrent neural networks have also gained immense traction as they have been shown to outperform all other state-of-the-art machine learning tools for handling very large dimensional data spaces. While most of the current applications involve image, video, speech and natural language processing, very recent studies have begun to explore its applicability for decision and control problems such as reinforcement learning~\cite{MKS15} and guided policy search~\cite{ZKLA15,LFDA15}. In this context, although supervised learning of a state to action map may be a straightforward formulation, we show that a similar performance can be achieved while learning the reward (VOI in this case) given the state and action combination via use of a deeper architecture. We compare the performance of the deep learning based VOI estimator with a more traditional, hand-tuned policy learner such as an augmented Markov Decision process (AMDP). A feasibility study was performed with a realistic human-machine information gathering scenario for a dynamic search problem.

%



\section{Human-Machine Collaboration}\label{sec:hmc}



\subsection{Problem Setup}
For concreteness, we will mainly focus throughout this work on information-gathering tasks involving localization of moving intruders (targets) in large environments that are incompletely covered by networks of mobile/static human and machine sensors. This serves as a useful analog for intelligence, surveillance and reconnaissance applications in the defense domain \cite{Kingston2012}, as well as safety and monitoring applications. For simplicity, we focus here on localizing a single known target in a 2D environment with a known map. We also focus here on a centralized sensor network architecture, in which all machine and human sensors report observations back to a single processing point. We further assume for now that a single human sensor is tasked to provide soft data observations and is always connected to the fusion center through an appropriate interface (e.g. a workstation console or mobile device).

At discrete time step $k$, let $X_k = X \in \RealSpace{2}$ be the unknown target location with initial prior probability distribution $p(X_0)$ at $k=0$. Assume that the environment can be modeled as a 2D grid with $n_g$ total elements, so that $X_k$ takes on discrete realizations $x^i_k$ for $i \in \set{1,...,n_g}$. A discrete time Markov chain with known transition probability $p(X_{k}|X_{k-1})$ is used to account for the target's uncertain motion through the environment. The Chapman-Kolmogorov equation dictates the evolution of $p(X_{k-1})$ to $p(X_{k})$,
\begin{align}
p(X_k) = \sum_{x^i_{k-1}} p(X_{k}|X_{k-1} = x^i_{k-1}) p(X_{k-1}=x^i_{k-1}).\nonumber
\end{align}
If hard sensor observations $O^h_k$ and soft sensor observations $O^s_k$ are available at each time step in the superset of observations ${\cal O}_{k} = \set{O^h_k, O^s_k}$, then the recursive Bayes' filter replaces $p(X_{k})$ with the conditional posterior distribution $p(X_{k}|{\cal O}_{1:k})$, where ${\cal O}_{1:k} = \set{{\cal O}_{1},\cdots,{\cal O}_{k}}$ and Bayes' rule gives
\begin{align}
&p(X_k|{\cal O}_{1:k}) \propto p(X_k|{\cal O}_{1:k-1}) p({\cal O}{k}|X_k), \nonumber \\
&\propto p(X_k|{\cal O}_{1:k-1}) p(O^h_k|X_k) p(O^s_k|X_k) \nonumber \\
&\propto \sum_{x^i_{k-1}}p(X_{k}|X_{k-1})p(X_{k-1}|{\cal O}_{1:k-1}) p(O^h_k|X_k) p(O^s_k|X_k), \nonumber
\end{align}
where $O^h_k$ and $O^s_k$ are assumed conditionally independent given the true target state. The posterior summarizes all information gathered by all sensors up to time $k$ and thus efficiently updates the belief in $X_k$ without requiring storage of ${\cal O}_{1:k}$. The information from each sensor is encapsulated by the observation likelihood functions $p(O^h_k|X_k)$ and $p(O^s_k|X_k)$. For hard sensors, $p(O^h_k|X_k)$ is typically derived from hardware specifications, or parametrically modeled and estimated via calibration. For soft sensors, however, $p(O^h_k|X_k)$ must be approximated as a function of $X_k$ depending on the type of human sensor interface used for a particular application. Conventional approximate Bayesian filtering techniques for non-Gaussian hard sensor data fusion (e.g. using grid, particle, or Gaussian mixture Kalman filter representations) can be naturally extended to incorporate fusion of soft data provided in various forms, including binary detection/no detection observations \cite{Ahmed-ICCPS-2015, Bourgault08} and semantic natural language observations \cite{AhmedTRO13, Park2013}. Suitable models $p(O^s_{k}|X_k)$ can be learned from data to properly account for uncertainty in human-generated information and capture key observation characteristics when deployed with real (expert or non-expert) humans. 

\begin{figure}[h]
 \centering
 \includegraphics[width=0.45\textwidth]{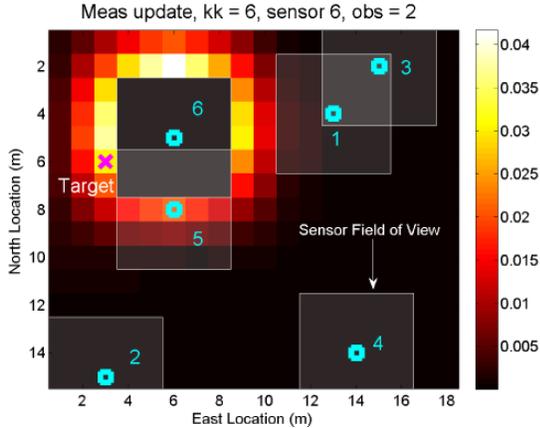}\vspace{-10pt}
\caption{\scriptsize Sample truth model simulation of 2D target localization problem, showing locations of cameras 1-6 and associated fields of views, along with posterior distribution $p(X_k|{\cal O}_{1:k})$ (heat map) and true target location (magenta x) for a random walk motion model $p(X_{k+1}|X_{k})$. Each sensor has $0.99$ detection rate and false alarm rate of $1 \times 10^{-4}$, 0.1504, 0.1585, 0.1992, 0.1510, and 0.1593, respectively.}
\label{fig:grid_example}
\vspace{-10pt}
\end{figure}
\vspace{-0pt}

The sensor scheduling problem in this context thus becomes one of obtaining an appropriate set of hard/soft measurements in the superset ${\cal O}_{1:k}$ so that the posterior distribution remains as informative as possible for constrained intelligent decision-making (e.g. deploying security assets to intercept the intruder within a certain time window). For example, consider the 2D grid world shown in Fig. \ref{fig:grid_example}, which features 6 cameras located in large open environment with a moving target that obeys random walk dynamics. Some of the cameras may be linked to automatic target recognition (ATR) algorithms, with known true detection and false alarm rates that define $p(O^h_k|X_k)$.  Alternatively, a remote human analyst can also access each camera in order to declare whether or not the target is in the camera's field of view (independently of ATR software). Due to the varying quality of each camera and various cognitive loads, the human analyst may incorrectly declare that the target has (not) been detected at any given camera with some true detection and false alarm rates defining $p(O^{s}_k|X_k)$. In either case, we assume the data fusion center can only request one hard observation $O^s_k$ or soft observation $O^h_k$ at any given $k$ from a single camera, due to bandwidth and processing restrictions. The scheduling problem thus addresses the question of which single har or soft sensor to query for the Bayes filter update at time $k$. This leads to the more general problem of obtaining an optimal sequence of hard/soft sensor queries to maximize some utility function over time. Matters become even more interesting and challenging when we consider that human sensors can also provide detailed semantic observations, e.g., `Target is now headed north towards the door very quickly'. In this case, given a known dictionary of grounded semantic statements that can be translated into target state information, the data fusion center could also schedule and execute the most informative semantic human sensor queries over time, which could be answered in free-form (e.g. `Tell me what you see in camera 2'?) or binary manner (`Is the target moving to the door in camera 2?'). 

\vspace{-10pt}
\subsection{Value of Information (VOI) for Soft/Hard Sensor Scheduling}
Previous work on combined soft/hard sensing focused primarily on instances where human sensors voluntarily `push' useful information as they see fit. However, this can lead to suboptimal gains for machine sensing and planning performance, especially if the human analyst becomes distracted or fails to recognize when machine sensors are unable to collect good data. This leads us to consider formal strategies for opportunistically `pulling' information from human sensors in the right way at the right time to enhance state estimation and long-term decision-making under uncertainty. Such interactions should also account for the costs of interacting with human sensors, in order to help manage their limited cognitive resources and avoid task overloading. 

These issues are naturally addressed via formal decision-theoretic Bayesian inference to assess the \emph{value of information} (VOI) for different soft data reports \cite{Kaupp2010}. For the simple target localization problem, suppose $O^s_k \in \set{o^{s,1}_k,...,o^{s,n_s}_k}$, where $o^{s,j}_k \in [0,1]$ denotes a specific kind of binary soft observation report. For instance, $o^{s,j}_k$ could represent a detection/no detection event for camera $j$, or a binary true/false response to a semantic query $j$ from a large list of possible queries. Given a utility function $U(D_k,X_k)$ representing the expected long-term benefit of taking some discrete action $D_k$ while the target is in state $X_k$, the VOI for receiving a single noisy report $o^{s,j}_k$ in response to a soft data query is

\begin{align}
&\mbox{VOI}(o^{s,j}_k) = \label{eq:VOIdef} \\
&\EV{\max_{D_k} U(D_k,X_k)}_{(o^{s,j}_k,X_k)} - \max_{D_k} \EV{U(D_k,X_k)}_{(X_k)}, \nonumber
\end{align}

where $\EV{f}_{(v)}$ is the expected value of $f$ over random variables $v$, and

\begin{align}
&\EV{\max_{D_k} U(D_k,X_k)}_{o^{s,j}_k,X_k} = \nonumber \\
&\sum_{o^{s,j}_k} p(o^{s,j}_k|{\cal O}_{1:k-1}) 
 \bigsquare{ \max_{D_k}  \sum_{x^i_k} p(X_k| {\cal O}_{1:k-1},  o^{s,j}_k) U(D_k,X_k)}, \nonumber \\
&\max_{D_k} \EV{U(D_k,X_k)}_{X_k} = \max_{D_k} \sum_{x^i_k} p(X_k| {\cal O}_{1:k-1}) U(D_k,X_k). \nonumber
\end{align}

Assuming a cost $c(o^{s,j}_k)$ for obtaining $o^{s,j}_k$, then the human sensor should be queried for $o^{s,j}_k$ if $\mbox{VOI}(o^{s,j}_k) > c(o^{s,j}_k)$. Thus, \pareqref{VOIdef} gives a formal way to assess whether the expected information from report $o^{s,j}_k$ is worth the cost of retrieving it, regardless of the outcome of $o^{s,j}_k$. The key idea in eq. \pareqref{VOIdef} is to compare the maximum expected utility given all possible outcomes for $X_k$ and $o^{s,j}_k$ to the maximum expected utility if no new soft data were obtained. In practical applications, we must compare the VOI at time $k$ for $n_s$ alternative sensor queries $o^{s,j}_k$, $j \in \set{1,...,n_s}$, and select only the query with the highest VOI. This is referred to as a \emph{myopic approximation}, since it does not consider all possible combinations of observations $o^{s,j}_k$ that could be taken together at time $k$ 
For dynamical systems, this approach is also myopic in the since it does not consider future sensing actions for time $k+1$ and beyond. However, the definition of VOI can be generalized to find an optimal non-myopic soft querying sequence $O^s_{k:k+T}$ for $T>0$, by assessing the single best query $O^s_k$ at each time step $k,...,k+T$, and comparing the final expected utility at step $k+T$ to the expected utility with respect to $p(X_{k+T}|X_{k},{\cal O}_{1:k})$ (the propagated belief without any future soft observations).

VOI depends heavily on $U(D_k,X_k)$ and $c(o^{s,j}_k)$, as well as the uncertainty in $p(X_k|{\cal O}_{1:k})$. For human sensors, $c(o^{s,j}_k)$ can be related to the expected cognitive cost of re-tasking human sensors \cite{Kaupp2010}. As there is no standard way to define $c(o^{s,j}_k)$ for general applications, for simplicity and without loss of generality, we ignore $c(o^{s,j}_k)$ for now and only consider utility defined by expected information gain for sensing actions. In this case, $D_k$ is related only to the choice of $j \in \set{1,...,n_s}$ and we seek to minimize the entropy of $p(X_k|{\cal O}_{1:k-1}, o^{s,j}_k)$, so that
\begin{align}
U(o^{s,j}_k,X_k) = \log p(X_k|{\cal O}_{1:k-1}, o^{s,j}_k).
\end{align}
Hence, the VOI for $o^{s,j}_k$ in \pareqref{VOIdef} becomes the expected decrease in posterior entropy,
\begin{align}
&\mbox{VOI}(o^{s,j}_k)  = \EV{ {\cal H}[p(x_k|{\cal O}_{1:k-1}, o^{s,j}_k)] }_{(o^{s,j}_k)} - {\cal H}[p(x_k|{\cal O}_{1:k-1})], \nonumber \\
&\mbox{where} \ {\cal H}[p(x_k|{\cal O}_{1:k})] = \EV{ \log p(x_k|{\cal O}_{1:k-1})  }_{(x_k)} \nonumber
\end{align}
This means that soft data $o^{s,j}_k$ will be (myopically) requested to keep the overall spread of uncertainty in $p(x_k|{\cal O}_{1:k})$ as small as possible. Entropy minimization is also widely used for tasking of hard sensors in target localization applications \cite{Huber08}, and thus provides a useful common objective for combined hard-soft sensor scheduling.

\subsubsection{Practical VOI inference and optimization}
Although VOI is an ideal measure by which to formally regulate human-machine interaction, VOI calculations are computationally expensive and lead to NP-hard Bayesian inference calculations for marginal observation likelihoods $p(o^{s,j}_k|{\cal O}_{1:k-1})$. The comparison of VOI for various $o^{s,j}_k$ reports can also be quite expensive when $n_s$ is large (e.g. for large semantic dictionaries), or if temporally non-myopic query sequences over multiple time steps are considered (due to combinatorial blow up). Hence, even for myopic approximations, it is generally impractical to explicitly compute and compare the VOI among $n_s$ soft sensing alternatives. While many approximate inference methods have been developed to address these issues \cite{Liu-UAI-2012}, these are still computationally expensive to implement as they still require online inference and optimization. 

It is worth noting that VOI-based sensor scheduling problems can also be interpreted as partially observable Markov decision processes (POMDPs) \cite{Krishnamurthy-TSP-2007}, which advantageously allow \emph{optimal sensing policies} $\pi(b(X_{k-1}))$ to be computed offline via value iteration over the belief space $b(X_{k-1}) = p(X_k|{\cal O}_{1:k-1})$. Such policies represent direct `look-up tables' from $b(X_{k-1})$ to sensing queries $o^{s,j}_k$ that automatically account for VOI through expected cumulative rewards, and thus allow for efficient online optimal querying that bypasses explicit computation and comparison of all possible future sensing actions.  Although there are many well-known approximate techniques for solving standard POMDPs with linear additive reward functions over $X_k$, information-based utilities such as negative entropy lead to non-standard POMDPs, since the expected rewards are non-linear functions of $b(X_{k-1})$. Hence, many state-of-the-art approximate POMDP solvers cannot be directly applied here. However, as discussed in Section 4, approximate methods such as the augmented Markov decision processes (AMDPs) \cite{Roy-JMLR-2005, Thrun-ProbRobbook-2005} can be used to derive sensing policies by first learning and then solving MDPs over \emph{belief space features} $f(b(X_k))$. AMDPs recover the linear additive reward structures in $f(b(X_k))$ space and thus enable the use of standard value iteration solutions for offline approximate policy generation for non-standard POMDPs. This is possible due to the fact that $b(k)$ is the sufficient statistic for POMDP policy calculations. However, it is non-trivial to define the features $f(b(k))$ which re-produce the optimal total expected rewards for the original POMDP. Furthermore, the corresponding MDP model parameters must be obtained via simulation for offline policy calculation, subject to hand-tuning of discount and immediate expected reward parameters.
%

Nevertheless, the AMDP approximation for POMDPs provides insight into other possible approximation strategies for optimal VOI-based sensor scheduling. In particular, it suggests that optimal querying policy maps could be obtained offline via supervised learning, using features of $b(k)$ and simulated observation instances to provide generalizable VOI associations between $b(k)$ and $O^{s}_k$ queries. The resulting learned `look up table' function could be trained for non-myopic querying, and would thus provide a computationally efficient means for pulling information from soft sensors that avoids expensive online brute force VOI optimization over all possible query sequences.

\begin{figure*}[htb]
 \centering
 \includegraphics[width=0.7\textwidth, trim={0cm 1cm 0cm 0cm}]{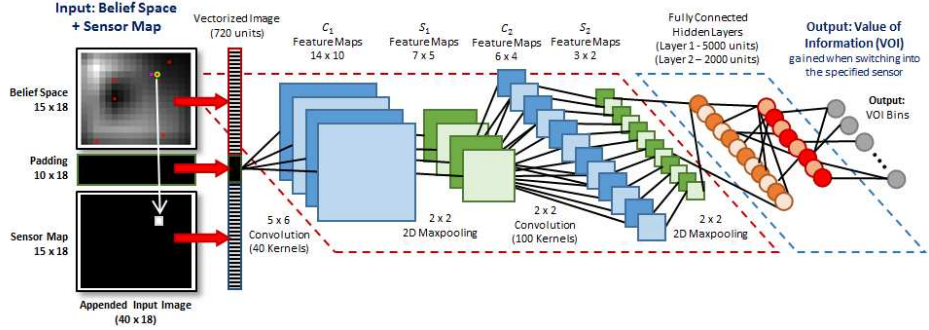} \caption{\scriptsize Schematic of the CNN used for reward learning.}
\label{fig:cnn_formulaC}\vspace{-5pt}
\end{figure*}

\section{Deep VOI Estimators}\label{sec:voi}
Deep neural networks present such an attractive option for estimating VOI without policy hand-tuning. Before describing the deep VOI estimation framework, we begin this section a brief background on deep Convolutional Neural Networks (CNN).

\subsection{Deep Convolutional Networks}\label{sec:CNN}
For the purpose of the study, deep CNN is a suitable choice due to its ease of training while still achieving a comparable performance despite the fact that it has fewer parameters relative to other fully connected networks with the same number of hidden layers. Furthermore, CNNs are designed to exploit the 2D structure of an input image (images of belief space in this case) by preserving the locality of features via the utilization of spatially-local correlations of an image through the use of tied weights, therefore being invariant to translations.

In CNNs, data is represented by multiple feature maps in each hidden layer. Feature maps are obtained by convolving the input image by multiple filters in the corresponding hidden layer. To further reduce the dimension of the data, these feature maps typically undergo non-linear downsampling with a $2 \times 2$ or $3 \times 3$ maxpooling. Maxpooling essentially partitions the input image into sets of non-overlapping rectangles and takes the maximum value for each partition as the output. After maxpooling, multiple dimension-reduced vector representations of the input are acquired and the process is repeated in the next layer to learn a higher representation of the data. At the final pooling layer, resultant outputs are linked with the fully connected layer where sigmoid outputs from the hidden units are joined with output units to infer a predicted class based on the highest joint probability given the input data.

The probability of an input vector $x$ being a member of the class $i$ can be written as:
\begin{equation}\label{eqn:prob_classification}
\text{Pr}(Y=i | \textbf{x},\textbf{W},\textbf{b}) = \text{softmax}_i (\textbf{Wx}+\textbf{b}) = \frac{e^{\textbf{W}_i \textbf{x} + \textbf{b}_i}}{\sum_j e^{\textbf{W}_j x + \textbf{b}_j}}
\end{equation}
\noindent and the prediction of the model is the class with the highest probability:
\begin{equation}\label{eqn:prob_ypred}
y_{\text{pred}} = \text{argmax}_i \text{Pr}(Y=i|\textbf{x},\textbf{W},\textbf{b})
\end{equation}
The model weights, $\textbf{W}$ and biases, $\textbf{b}$ are optimized by an error backpropagation algorithm, where true class labels are compared against the model prediction using an error metric that becomes the loss function of the algorithm. Specifically, the loss function $\ell$ to be minimized for a dataset $\textbf{X}$, parametrized by $\theta$ is:
\begin{equation}\label{eqn:nll}
\ell(\theta=\{\textbf{W},\textbf{b}\},\textbf{X}) = - \sum_{i=0}^{\mathcal{|\textbf{X}|}} \left[ \log{\left(\text{Pr}(Y=y^{(i)}|x^{(i)},\textbf{W},\textbf{b})\right)} \right]
\end{equation}
\noindent where $y^{(i)}$ denotes the class index.

\subsection{Design of framework}\label{sec:framework}

\vspace{5pt}
\textbf{Policy Learning:} 
A typical formulation of the problem is directly mapping the belief space to the action by using belief maps as inputs and sensor indices as outputs. For this problem set-up, a single belief map is accompanied by an index denoting the action of switching into the sensor that gives the best VOI gain. The belief map is represented by a $15\times 18$ (rows $\times$ columns, in pixels) grayscale image, where pixels with high grayscale intensity denote high belief and pixels with low grayscale intensity denote low belief. The belief map is then vectorized into a vector of $1\times 270$ units and goes through a typical convolutional neural network to make a prediction that suggests which sensor to switch into for the most VOI gain.



\vspace{5pt}
\textbf{Reward Learning:} 
Another formulation of the problem that is perhaps more robust, we aim to learn the underlying function for predicting VOIs given a belief map and an arbitrary sensor index. For this formulation, the CNN architecture is similar but the inputs are modified to include an arbitrary sensor. The Cartesian coordinates of this sensor is indicated on a map with the same dimensions of the belief map called the \textit{sensor map}. In addition, enough padding is added between the belief map and the sensor map such that the information from these two spaces do not interfere with the learning of filter weights. The belief map $(15\times 18)$, padding $(10\times 18)$, and sensor map $(15\times 18)$ are vertically concatenated to form a $40\times 18$ image that will become the input to the CNN. VOI also becomes the true class labels associated with this input image during training. However, since VOI is real-valued, we have discretized the VOI space into 62 classes to formulate a classification problem. Hence, the outputs of the model (i.e., the classes) correspond to the estimated VOI given a sensor and a belief map. Fig.~\ref{fig:cnn_formulaC} shows a schematic of this formulation.


Using Eq.~(\ref{eqn:prob_classification}), (\ref{eqn:prob_ypred}) and (\ref{eqn:nll}), the VOI can be estimated based on the associated action of switching into a particular sensor given the current belief map.

\vspace{5pt}
\textbf{Training Data Generation:}
Data is generated from running the simulation for 10,000 time steps with random sensor false alarm rate of $15\pm5\%$ and the target moving in a random walk manner. Depending on the formulation of the problem, the resultant size of the dataset varies. In the policy learning formulation, we are limited to one training example per time step since there is only one optimal sensor index given a single belief map. This results in 10,000 examples, where half of them (i.e., 5,000) is used for training and the other half is used for cross-validation to avoid overfitting. The reward learning formulation allows us to produce six examples per time step (since there are six sensors in the field) and generated a total of 60,000 examples. In this case, we consider the expectation of VOI gained 4 time steps ahead in future, in a non-myopic fashion. The expected VOI gained by switching into the particular sensor is used as the class labels after discretization. Again, half of the example data is used for training and the other half is used for validation.

\textbf{Network Architecture and Hyper-parameters:}
Specific details on the number of filters, convolutional layer, pooling layers, and fully connected layers are shown in 
and Fig.~\ref{fig:cnn_formulaC}. A learning rate of 0.01 is used for the gradient descent algorithm for training the CNN in a supervised manner with a batch size of 10 samples. The optimized model is acquired prior to the point when validation error becomes consistently higher than the training error in subsequent training iterations.

\vspace{-10pt}
\section{Results and Discussion}\label{sec:results}

In this section, the performance of CNN model is evaluated and compared with AMDP.

\subsection{Prediction Accuracy}\label{sec:performance}

\begin{figure}[!htb] 
 \centering
 \includegraphics[width=0.42\textwidth, trim={0cm 1cm 0cm 1cm}]{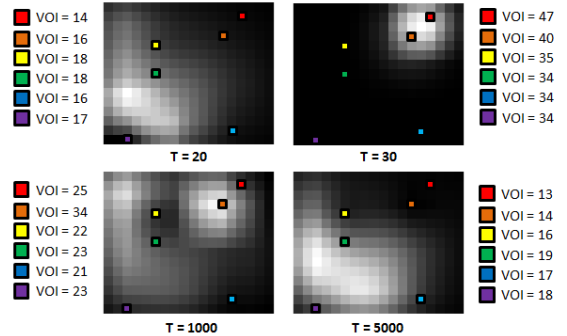} \caption{\scriptsize Four examples of belief map and VOI (discretized into bins) associated with the action of switching into a particular sensor at different simulation time steps $T$.}
\label{fig:voi_example}\vspace{-10pt}
\end{figure}


\begin{figure}[h] 
 \centering
 \includegraphics[width=0.35\textwidth]{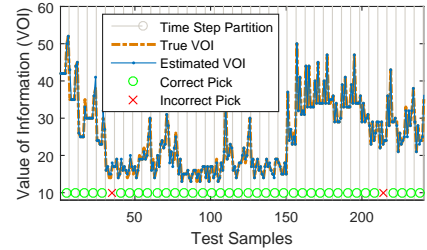}\vspace{-10pt} \caption{\scriptsize Estimated VOI vs. True VOI for the first 40 out of 10,000 simulation time steps. The figure also shows whether the sensor selected through rank-ordering the predicted VOI is the same as the sensor selected by ordering the true VOI, represented by green circles (matching) and red crosses (non-matching).}
\label{fig:timesteppred_big}\vspace{-25pt}
\end{figure}

The VOI estimated by the trained model follows closely to the true VOI computed as shown in Fig.~\ref{fig:timesteppred_big}. Each point in one time step partition represents a sensor index; the model has to select one sensor out of six based on the current belief map with the highest expected VOI gain. A successful selection is represented by a green circle, whereas a wrong selection is represented by a red cross. It is observed that CNN can be in fact trained to learn the reward function and produce VOIs similar to the values computed by the brute force simulation. Most importantly, the high capability of predicting this non-myopic VOI suggests that the model does not naively generate VOI that is directly proportional to the grayscale pixel intensity of the belief map (which corresponds to the probability of finding the target) at that particular sensor location. Furthermore, careful inspection of the camera choice errors reveal that most of the errors come from confusing between camera 5 and camera 6 that are spatially very close as well as tend to have similar VOI reward for many belief space configurations. 

\begin{figure}[!htb] 
 \centering
 \includegraphics[width=0.35\textwidth]{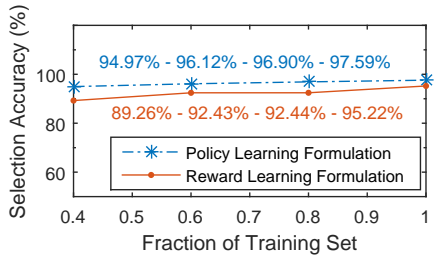}\vspace{-10pt} \caption{\scriptsize Degradation of selection accuracy with reduction of dataset sizes used to train the CNN model.}
\label{fig:degrade}\vspace{-7pt}
\end{figure}

Results also suggest that the trained CNN model is able to generalize into unseen belief maps and still predict VOI that is close to the true VOI. To show this, we have reduced the size of the dataset used to train the CNN model with $p\in\{0.4,0.6,0.8,1.0\}$ denoting the fraction of the new training and validation set sizes from their original sizes. From Fig.~\ref{fig:degrade}, the accuracy of selecting the best sensor with largest VOI gain decreases slightly with smaller dataset sizes. Policy learning formulation (i.e., direct action mapping) seems to perform slightly better compared to reward learning (i.e., mapping into VOI) as it suffers only a slight degradation in prediction performance. Furthermore, policy learning is attractive because it is straightforward and intuitive - given a belief map, a decision can be made to choose the best sensor. Note, the reward learning framework needs a deeper architecture (i.e., an extra fully connected layer) to achieve a similar performance compared to the policy learning framework. Training the CNN for policy is also a little easier as it allows a simpler model with the number of output classes equivalent to the number of available options (i.e., choosing a particular sensor). On the other hand, the ability to evaluate the VOI from an action carried out in a specific belief space is very useful in situations where the location/charecteristics of the sensors may be altered slightly. Additionally, the framework can be more flexible in terms of additional objectives or imposing extra constraints (e.g., the predicted VOI can be processed to include penalization terms for certain actions). If this happens, the policy learning model must be retrained whereas the reward learning model may not need to be retrained. The adaptability of the reward learning formulation is absolutely valuable for generalizing into larger problem setups by paying a small price in accuracy. 

\subsection{Comparison with AMDP}\label{sec:amdp}
A policy derived from a feature-based solution to the POMDP model for the sensor scheduling problem can be used to provide a baseline comparison with the learned CNN policies. As mentioned in Section 2, augmented Markov Decision processes (AMDPs) can be used to solve the non-standard POMDP for sensor scheduling when the reward function is defined to minimize the entropy of the posterior state belief $b_k = p(X_k|{\cal O}_{1:k})$. AMDPs use a learned Markov decision process (MDP) model on features $f(b_k)$ of the belief space, which can include the entropy of $b_k$. This MDP can then be used to derive scheduling policies via offline value iteration with linear additive rewards defined in terms of $f(b_k)$ instead of $X_k$. By choosing a good set of features, the optimal expected total reward for the AMDP policy can closely match that of the original POMDP defined over $X_k$ with non-standard entropy rewards. As discussed in \cite{Roy-JMLR-2005, Thrun-ProbRobbook-2005}, the key idea behind the AMDP is that most of the reachable belief space $b_k$ evolves along a low-dimensional manifold, which can be encoded by a feature set whose size is typically much smaller than the number of possible states $X_k$ in the original POMDP. Hence, AMDPs offer a generative feature-based alternative to finding scheduling policies, in contrast to the discriminative feature-based modeling approach used by CNNs.

However, the problem remains to find suitable features $f(b_k)$ and learn the corresponding MDP model over the feature space, which is a non-trivial learning problem. Furthermore, discount and immediate expected reward parameters for the AMDP must be hand-tuned to arrive at suitable policies via value iteration. Finally, stationary infinite horizon policies must often be used in practice to avoid the computational cost of performing finite time value iterations for non-stationary scheduling policies (especially if discount and reward parameters must be tuned). The use of infinite horizon policies necessarily makes the AMDP suboptimal for finite-horizon querying problems, but nevertheless provides insight into the expected capabilities of generative feature-based learning approaches for sensor scheduling policy estimation. Our implementation of AMDP for the simple 2D grid world problem follows the basic technique presented in \cite{Thrun-ProbRobbook-2005}, which defines $f(b_k)$ as the stacked vector of the maximum a posteriori (MAP) state $X_k$ of $b(k)$ and the (discretized) entropy of $b(k)$. The resulting implementation used 27,000 AMDP feature states, and the policy was generated using hand-tuned rewards (with positive rewards proportional to inverse square of entropy for transitions to lower-entropy states, and negative rewards otherwise) and a fixed value function discount factor of 0.1 (to encourage earlier transitions to low-entropy states). 
\begin{figure}[h]
 \centering
 \includegraphics[width=0.47\textwidth]{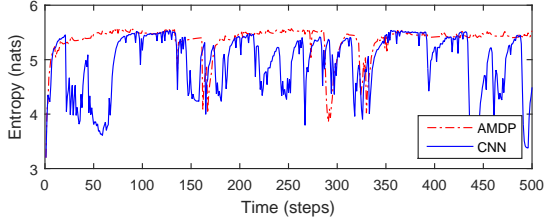}\vspace{-5pt} \caption{\scriptsize Entropy of belief map over time.}
\label{fig:entropy}\vspace{-5pt}
\end{figure}

To study the level of certainty where the target is in a particular region, the entropy of the belief map at each time step is studied. High entropy means there is a large uncertainty that the target is in a given region, and low entropy suggests an ability of narrowing down the target location. The entropy of the belief map generated using AMDP and CNN is computed and plotted over 500 time steps, shown in Fig.~\ref{fig:entropy}. Belief entropy from AMDP experiences sharp dips in magnitude occasionally but stays high at all other times. Entropy from CNN is generally lower at most time steps, thus implying lower uncertainty. However, this may not be so simple to say, especially given the target's propensity to move around a lot--even if the model starts off knowing exactly where the target is, the entropy jumps up a lot at the next time step. Therefore, along with this decision uncertainty metric, a correctness metric is also used to compare the performances of AMDP and CNN.

\begin{figure}[h]
 \centering
 \includegraphics[width=0.47\textwidth]{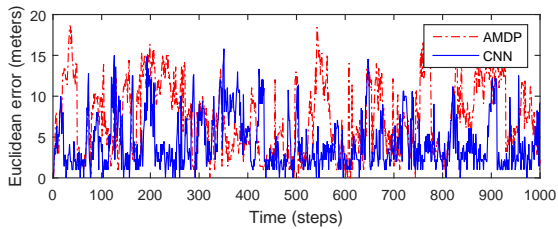}\vspace{-5pt} \caption{\scriptsize Euclidean error between MAP estimate and target location over time.}
\label{fig:modeerrortime}\vspace{-10pt}
\end{figure}

\begin{figure}[h]
 \centering
 \includegraphics[width=0.4\textwidth]{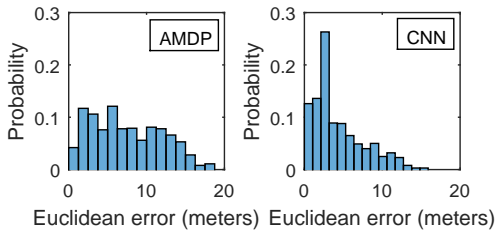}\vspace{-5pt} \caption{\scriptsize Error distribution between MAP estimate and target location.}
\label{fig:disterrordistribution}\vspace{-10pt}
\end{figure}

Estimation error is defined as the Euclidean distance between the maximum a posteriori probability (MAP) estimate (i.e., largest value of the belief map) and the actual target location. Ideally, this error should be as close to zero as possible. The error is calculated using belief map and target locations for each time step (see Fig.~\ref{fig:modeerrortime}) and its distribution is shown in Fig.~\ref{fig:disterrordistribution}. Clearly, CNN outperforms AMDP as the distribution is more skewed with higher probability of the belief map mode being closer to the actual target location, hence increasing the success rate of target-tracking.

\begin{rem}
As a hierarchical feature extraction tool, deep CNN is able to extract belief space features at different spatial scales as well as obtain `features of features'. Also, belief spaces generally evolve on low-dimensional manifolds (i.e., feature spaces) as suggested by the AMDP formulation. Therefore, it becomes feasible for a deep CNN to associate these representative features to the VOI. In this context, we make an interesting observation (as shown in Fig.~\ref{fig:degrade}) that CNN performance
does not degrade much with a drastic decrease in the training data size. This probably suggests that there is a relatively low number of key features in the belief space that (the CNN is able to learn) lead to various belief space configurations via complex combinations. Another view could be that the proposed framework aims to automatically learn the probability density of the reward given belief space and action. Therefore, with a hierarchy of nonlinear functions and a large number of model parameters, it becomes feasible to model arbitrarily complex densities. However, sufficient training data and regularization steps are necessary to avoid overfitting.
\end{rem}

\section{Semantic Soft Data Scheduling}\label{sec:real}
Consider a `cops and robbers' scenario in which one robot (the `cop') searches for another mobile intruder robot (the `robber') in an indoor environment, with a simulated remote human `security guard' providing observations of the robber's 2-dimensional position state. As in the previous grid world toy problem, the cop maintains a Bayesian belief map over the environment of the robber's state, and updates this belief either through fusion of hard sensor measurements (e.g. detection of robber position via an on-board camera) or soft data. The soft data in this case takes the form of a human semantic observation as in \cite{AhmedTRO13}, e.g. ``The robber is in front of the desk.'' We assume that the remote human views the scene either through the cop robot's camera, or a security camera placed in each room. The human thus has full visibility into the space, but is constrained to visibility of no more than one room per time step. Similarly, we have constrained the cop robot to asking a single question about the robber once every $n_q =5$ time steps, as a way to mitigate operator load. Over time, the probability mass of the target position diffuses over the environment, based on the cop's estimate of the robber's position and known random-walk dynamics model.

\begin{figure}[!htb]
 \centering
 \includegraphics[width=0.45\textwidth]{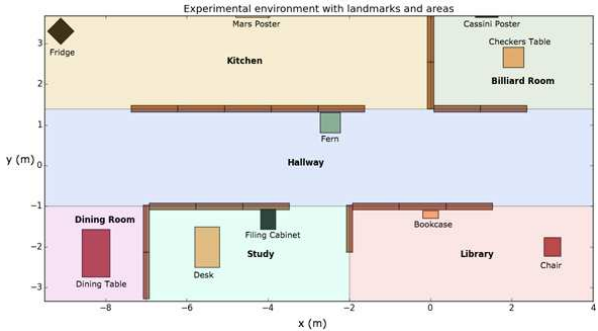}\vspace{-5pt} \caption{\scriptsize The indoor search environment and associated semantic features for soft observations.}
\label{fig:fleming_environment}\vspace{-10pt}
\end{figure}

\begin{figure}[!htb]
 \centering
 \includegraphics[width=0.35\textwidth]{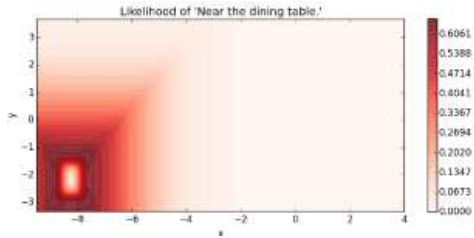}\vspace{-5pt} \caption{\scriptsize The likelihood function that maps to the example observation, `Target is near the dining table'.}
\label{fig:near_likelihood}\vspace{-5pt}
\end{figure}

With this setup, the human observations $o^{s,j}_k$ can be thought as binary `true/false' responses to the questions asked by the cop robot, where $j$ indexes an element from a finite list of $n_s$ semantic questions. We assume $o^{s,j}_k$ arrive only as responses to questions posed by the robot, i.e. the human does not `push' information voluntarily, and only one semantic query $j$ is selected from $\set{1,...,n_s}$. The cop can ask the human whether the robber is either inside one of the rooms or near one of the objects shown in Fig. \ref{fig:fleming_environment}. This leads to $n_s=16$ possible questions that generate measurement updates. For example, if the cop asks, "Is the robber near the dining table?", a positive response triggers fusion of the likelihood shown in Fig. \ref{fig:near_likelihood} with the current belief state. At any given query instance, the cop seeks to ask the question with maximum VOI, as defined by the expected reduction in entropy. The $16^T$ queries of depth $T$ are ranked by their corresponding VOI; in the non-myopic case of $T\geq 2$, all question paths are ranked by VOI first and then reduced to the initial $16$ possible questions, ranked by the maximum VOI of all query sequences starting with that question.

\begin{figure}[!htb]
 \centering
 \includegraphics[width=0.5\textwidth]{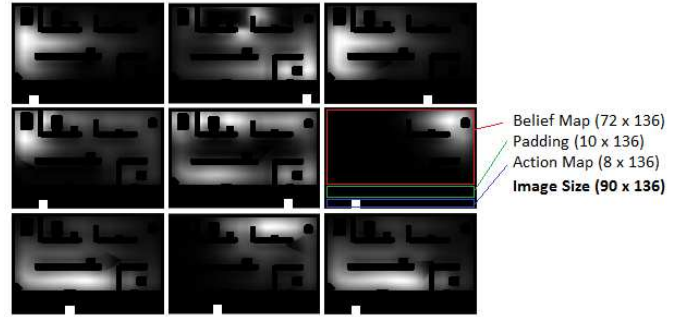} \caption{\scriptsize Nine examples of CNN input showing the belief map, padding, and the action map in the realistic scenario.}
\label{fig:realscene_input}\vspace{-5pt}
\end{figure}

\textbf{CNN training:} This scenario provides a truth model used to train the CNN. Applying the same CNN input-output structure and problem formulation, the belief map ($72\times136$) is appended with padding ($10 \times 136$) and the action map ($8\times 136$) where the action map is divided into 16 equally spaced nodes that light up depending on the question posed by the cop. Enough padding is created to separate the belief map from the action map for efficient learning of the convolving filters. Fig.~\ref{fig:realscene_input} illustrates the images used in learning. We trained the CNN model (only for myopic scenario for this feasibility study) with 31,680 training examples and 31,680 validation examples sampled from 100 simulation trials with 400 time steps each. Therefore, variability in the belief space is sufficient. As the input image size is now larger relative to the simpler problem, filter sizes are increased to 10 of $7\times9$ in the first convolutional layer and 25 of $5\times5$ in the second convolutional layer with all other parameters equal. Note that the length of the smaller padding edge (i.e., 10 pixels) is still larger than the longest filter dimensions.

\textbf{Results:} From our simulations, the CNN is \textbf{91.33\%} accurate in determining the next best question with highest VOI gain that should be posed by the robot (based on 1,980 time steps), as compared with the ground truth determined by brute force calculation. Interestingly, most of the errors occur as the target moves from one room to a neighboring one (e.g., from library to study) and the query selection sometimes remain associated with the previous room. The CNN-based 'query' selection quickly adjusts within a few time steps after such a transition. Hence, most of these errors can potentially be avoided with a non-myopic VOI implementation for this real-life scenario and is currently being pursued. Overall, this exercise demonstrates the feasibility of using deep learning architectures for such decision-making processes.


\section{Conclusions and Future Work}\label{sec:con}
VOI-based human-machine interfaces can automatically determine how and when to present queries to human operators in a real problem.
However, practical online implementation of VOI-based querying strategies remains challenging, since the problem of selecting the optimal sequence of queries leads to a difficult analytically intractable joint optimization and inference problem. This paper uses recent advancements in deep learning to build a VOI estimation framework that is shown to be able to reliably estimate VOI without any policy hand-tuning. 
A 2-D grid world search problem (with a moving target) is used to compare the performance of the finite horizon deep VOI estimator with that of a hand-tuned AMDP policy. Simulation results show that a CNN-in-the-loop information gathering system is able to lower the expected entropy of belief spaces and the expected error between the MAP estimate of the target and the true target compared to an AMDP-in-the-loop process. Finally, a feasibility study was performed on a simulation test bed with a realistic human-machine collaboration problem.

Better network design and hyper-parameter optimization are currently being investigated. Other future research directions are: (i) online tuning of deep networks; (ii) hybrid approaches involving deep feature extractors and traditional planning algorithms (e.g. using CNNs to learn feature maps for AMDP-based policy approximations); (iii) addressing sensor modeling issues, including potentially unknown false alarms/missed detection rates and imperfect human sensor models (e.g. using hierarchical Bayesian modeling as in \cite{Ahmed-ICCPS-2015} to account for model uncertainties); and (iv) validation of the proposed methodology on indoor robotic target search test bed with live human users. We will also extend the comparisons made here between feature-based direct policy learning approaches and feature-based POMDP approximations to other state-of-the-art POMDP approximations, including those that approximate low-dimensional reachable belief spaces via online sampling rather than through offline-learned feature compression \cite{Silver-NIPS-2010}.


%
\bibliographystyle{abbrv}
\bibliography{iccps2016}  
%
%


\end{document}

%% file: Ahmed_defs.tex
\newcommand{\EV}[1]{\mathbb{E}\left[#1 \right]} 
\newcommand{\bigsquare}[1]{\left[ #1 \right]}

\newcommand{\set}[1]{\left\{ #1 \right\}}
\newcommand{\RealSpace}[1]{\mathbb{R}^{#1}}

\newcommand{\pareqref}[1]{(\ref{eq:#1})}





%% file: main.bbl
\begin{thebibliography}{10}

\bibitem{Ahmed-ICCPS-2015}
N.~Ahmed, M.~Campbell, D.~Casbeer, Y.~Cao, and D.~Kingston.
\newblock Fully bayesian learning and spatial reasoning with flexible human
  sensor networks.
\newblock In {\em Proceedings of the ACM/IEEE Sixth Int'l Conf. on
  Cyber-Physical Systems}, pages 80--89. ACM, 2015.

\bibitem{AhmedTRO13}
N.~Ahmed, E.~Sample, and M.~Campbell.
\newblock Bayesian multicategorical soft data fusion for human robot
  collaboration.
\newblock {\em IEEE Trans. on Robotics}, 29:189--206, 2013.

\bibitem{Bourgault08}
F.~Bourgault, A.~Chokshi, J.~Wang, D.~Shah, J.~Schoenberg, R.~Iyer, F.~Cedano,
  and M.~Campbell.
\newblock Scalable {Bayesian} human robot cooperation in mobile sensor
  networks.
\newblock In {\em Int'l Conf. on Intelligent Robots and Sys.}, pages
  2342--2349, 2008.

\bibitem{Huber08}
M.~F. Huber, T.~Bailey, H.~Durrant-Whyte, and U.~D. Hanebeck.
\newblock {On entropy approximation for Gaussian mixture random vectors}.
\newblock In {\em 2008 IEEE Int'l Conf. on Multisensor Fusion and Integration},
  pages 181--188, Aug. 2008.

\bibitem{Kaupp-JFR-2007}
T.~Kaupp, B.~Douillard, F.~Ramos, A.~Makarenko, and B.~Upcroft.
\newblock Shared environment representation for a human robot team performing
  information fusion.
\newblock {\em Journal of Field Robotics}, 24(11):911--942, 2007.

\bibitem{Kaupp2010}
T.~Kaupp, A.~Makarenko, and H.~Durrant-Whyte.
\newblock Human robot communication for collaborative decision making: A
  probabilistic approach.
\newblock {\em Robotics and Autonomous Systems}, 58(5):444--456, May 2010.

\bibitem{KSB10a}
K.~Kavukcuoglu, P.~Sermanet, Y.-L. Boureau, K.~Gregor, M.~Mathieu, and
  Y.~LeCun.
\newblock Imagenet classification with deep convolutional neural networks.
\newblock {\em Neural Information Processing Systems}, 2010.

\bibitem{Kingston2012}
D.~Kingston.
\newblock {Intruder Tracking Using UAV Teams and Ground Sensor Networks}.
\newblock In {\em German Aviation and Aerospace Congress (DLRK 2012)}, Berlin,
  Germany, 2012. German Society for Aeronautics and Astronautics (DGLR).

\bibitem{Krishnamurthy-TSP-2007}
V.~Krishnamurthy and D.~V. Djonin.
\newblock Structured threshold policies for dynamic sensor scheduling: A
  partially observed markov decision process approach.
\newblock {\em IEEE Transactions on Signal Processing}, 55(10):4938--4957, Oct.
  2007.

\bibitem{LFDA15}
S.~Levine, C.~Finn, T.~Darrell, and P.~Abbeel.
\newblock End-to-end training of deep visuomotor policies.
\newblock {\em arXiv preprint arXiv:1504.00702}, 2015.

\bibitem{Liu-UAI-2012}
Q.~Liu and A.~Ihler.
\newblock {Belief Propagation for Structured Decision Making}.
\newblock In {\em Proceedings of Uncertainty in Artificial Intelligence (UAI)},
  2012.

\bibitem{MKS15}
V.~Mnih, K.~Kavukcuoglu, D.~Silver, A.~A. Rusu, J.~Veness, et~al.
\newblock Human-level control through deep reinforcement learning.
\newblock {\em Nature}, 518(7540):529--533, 02 2015.

\bibitem{Park2013}
B.~Park, A.~Johannson, and D.~Nicholson.
\newblock Crowdsourcing soft data for improved urban situation assessment.
\newblock In {\em Int'l Conf. on Information Fusion (FUSION)}, pages 669--675.
  IEEE, 2013.

\bibitem{Roy-JMLR-2005}
N.~Roy, G.~Gordon, and S.~Thrun.
\newblock {Finding Approximate POMDP Solutions Through Belief Compression}.
\newblock {\em Journal of Machine Learning Research}, 23:1--40, 2005.

\bibitem{Silver-NIPS-2010}
D.~Silver and J.~Veness.
\newblock {Monte-Carlo planning in large POMDPs}.
\newblock In {\em Advances in Neural Information Processing Systems}, pages
  1--9, 2010.

\bibitem{Thrun-ProbRobbook-2005}
S.~Thrun, W.~Burgard, and D.~Fox.
\newblock {\em {Probabilistic Robotics}}.
\newblock MIT Press, Cambridge, MA, 2005.

\bibitem{ZKLA15}
T.~Zhang, G.~Kahn, S.~Levine, and P.~Abbeel.
\newblock Learning deep control policies for autonomous aerial vehicles with
  mpc-guided policy search.
\newblock {\em arXiv preprint arXiv:1509.06791}, 2015.

\end{thebibliography}
